\documentclass[printer]{aa} 

\usepackage{graphicx}
\usepackage{amsmath}
\usepackage{txfonts}
\usepackage{csquotes}
\MakeOuterQuote{"}

\begin{document}

\title{Local ISM  3D distribution and soft X-ray background}
\subtitle{Inferences on nearby hot gas \bf{and the North Polar Spur}}

\author{ L. Puspitarini
\inst{1}
\and
R. Lallement
\inst{1}
\and
J.-L. Vergely
\inst{2}
\and
S. L. Snowden
\inst{3}
}

\institute{GEPI Observatoire de Paris, CNRS, Universit\'e Paris Diderot, Place Jules Janssen  92190 Meudon, France\\
\email{lucky.puspitarini@obspm.fr; rosine.lallement@obspm.fr}
\and
ACRI-ST, 260 route du Pin Montard, Sophia Antipolis, France\\
\email{jeanluc.vergely@latmos.ipsl.fr}
\and
Code 662, NASA/Goddard Space Flight Center, Greenbelt, MD 20771, USA\\ 
\email{steven.l.snowden@nasa.gov}
}

\date{}

\abstract
{Three-dimensional (3D) interstellar medium (ISM) maps can be used to locate not only interstellar (IS) clouds, but also IS bubbles between the clouds that are blown by stellar winds and supernovae, and are filled by hot gas. 
To demonstrate this, and to derive a clearer picture of the local ISM, we compare our recent 3D  maps of the IS dust distribution to the ROSAT diffuse X-ray background maps after removal of heliospheric emission.
In the Galactic plane, there is a good correspondence between the locations and extents of the mapped nearby cavities and the soft (0.25 keV) background emission distribution, showing that most of these nearby cavities contribute to this soft X-ray emission.
Assuming a constant dust to gas ratio and homogeneous 10$^{6}$ K hot gas filling the cavities, we modeled in a simple way the 0.25 keV surface brightness along the Galactic plane as seen from the Sun, taking into account the absorption by the mapped clouds. The data-model comparison favors the existence of hot gas in the solar neighborhood, the so-called Local Bubble (LB). The inferred average mean pressure in the local cavities is found to be on the order of $\sim$10,000 cm$^{-3}$K, in agreement with previous studies, providing a validation test for the method. 
On the other hand, the model overestimates the emission from the huge cavities located in the third quadrant. Using CaII absorption data, we show that 
the dust to CaII ratio is very small in this region, implying the presence of a large quantity of lower temperature (non-X-ray emitting) ionized gas and as a consequence a reduction of the volume filled by hot gas, explaining at least part of the discrepancy.  In the meridian plane, the two main brightness  enhancements coincide well with the LB's most elongated parts and \textit{chimneys} connecting the LB to the halo, but no particular nearby cavity is found towards the enhancement in the direction of the bright North Polar Spur (NPS) at high latitude.  
We searched in the 3D maps for the source regions of the higher energy (0.75 keV) enhancements in the fourth and first quadrants. Tunnels and  cavities are found to coincide with the main bright areas, however no tunnel nor cavity is found to match the low-latitude $b \gtrsim 8^{\circ}$, brightest part of the NPS. 
In addition, the comparison between the 3D maps and published spectral data do not favor the nearby cavities located within $\sim$ 200pc as potential source regions for the NPS. 
Those examples illustrate the potential use of more detailed 3D distributions of the nearby ISM for the interpretation of the diffuse soft X-ray background.}


\maketitle


\section{Introduction}

Multi-wavelength observations of interstellar matter or interstellar medium (ISM) in emission are providing increasingly detailed maps, bringing information on all phases. However, the information on the distance to the emitting sources and their depth is often uncertain or lacking. This applies in particular to the diffuse soft X-ray background (SXRB), an emission produced by $10^6$ K gas filling the cavities blown by stellar winds and supernovae, by the Galactic halo, and in a minor way by extragalactic sources. 

The R{\"o}ntgen satellite (ROSAT) was launched in 1990 and produced observations of the diffuse X-ray background in several bands, from 0.25 keV to 1.5 keV \citep{1995ApJ...454..643S,snowden97}, providing unique information on the nearby and more distant, soft  X-ray emitting hot  gas. However, the determination of the properties of this gas is not straightforward, because it depends on the knowledge of the ISM distribution, for both the X-ray emitting gas (dimensions of the volumes filled by hot gas) and the X-ray absorbing gas (foreground clouds). A number of developments have addressed this problem and succeeded in disentangling the various contributions to the signal by modeling the intensity and the broad-band spectra \citep{snowden97}, or using shadowing techniques based on IR maps \citep{1998ApJ...493..715S,2000ApJS..128..171S}. Information was thus obtained on the contributions to the diffuse background of extragalactic emission, emission from the bulge, from the halo, and from the nearby cavities. In particular, the ROSAT soft X-ray data (0.25 keV) brought unique information on the hot gas filling the so-called Local Bubble, a low density, irregular volume in the solar neighborhood. 

Still, many uncertainties remain in the physical properties of the emitting gas. As a matter of fact,  even if its temperature can be determined spectrally,  its density (or its pressure) can be derived only if one knows the extent of the emitting region along each line-of-sight (LOS), and the situation becomes rapidly very complex if several regions contribute to the signal. Finally, a strong additional difficulty is the existence of a contaminating foreground emission due to charge-exchange of solar wind ions with interstellar and geocoronal neutrals (the so-called SWCX emission). This signal was not understood at the time of the first ROSAT analyses, but is now increasingly well documented  (e.g., \cite{2004ApJ...607..596W, 2004ApJ...610.1182S, 2007PASJ...59S.133F}). It is highly variable with time and also depends on the observation geometry (look direction and line of sight through Earth's exosphere and magnetosheath). During the ROSAT All-Sky Survey (RASS), it manifested as sporadic increases of the count rate lasting for few hours to days (called long term enhancements, LTEs). Those LTE's were found to be be correlated with the solar activity, especially solar wind events \citep{1994PhDT.......103F}. \cite{2000ApJ...532L.153C} convincingly demonstrated that LTEs are the product of the highly ionized solar wind species acquiring electrons in an excited state from exospheric or heliospheric neutrals, which then emit X-rays.

The exospheric (geocoronal) contribution is highly variable with time and can be removed relatively easily, at least to a minimal zero level. This minimum zero level of geocoronal SWCX was not subtracted from the RASS data as it was not understood at the time.  The strong positional correlation and near-zero offsets of the $\frac{1}{4}$ keV data (Snowden et al. 1995) with the previous all-sky surveys of Wisconsin \citep{mccammon83}, SAS-3 \citep{marshall84}, and HEAO-1 A2 \citep{garmire92} suggest a minimal zero-level offset in the RASS data.  A more recent analysis of the correlation between the solar wind flux and LTE count rate suggests that the zero-level contribution of the geocoronal SWCX may be roughly 25\% of the minimum 1/4 keV band surface brightness found in the Galactic plane. 

The heliospheric SWCX emission is produced by interactions of solar wind ions with IS neutrals drifting through the solar system. This emission did not contribute significantly to the LTEs as the time scales are typically significantly longer than the hours-long to day-long variation of the geocoronal SWCX.  However the zero level of heliospheric SWCX is thought to be on the order of the 1/4 keV background itself and must be taken into account. Because the distribution of interstellar hydrogen and helium atoms in the heliosphere is governed by the relative motion between the Sun and the local interstellar cloud and is characterized by marked maxima along the corresponding direction, this SWCX contribution is expected to be oriented in the same way and have two strong maxima in opposite directions, one for hydrogen and one for helium. 
\cite{lall04} estimated the SWCX heliospheric contribution based on the hydrogen and helium atom distributions and, for the ROSAT geometry, found that parallax effects attenuate strongly the effects of these emissivity maxima and make the SWCX signal nearly isotropic for those ROSAT observing conditions. This has the negative consequence of making its disentangling more difficult and provides an offset in the measurement of the 0.25 keV background. This is unfortunate since, as said above, the heliospheric emission is far from negligible compared to the LB or halo emission, which adds to the complexity of the modeling, especially due to its anisotropy, temporal and spectral variability \citep{koutroumpa07,koutroumpa09a,koutroumpa09b} at long timescales.  The contribution of the heliospheric SWCX is explicitly considered in the following analysis.

At higher energies (0.75 keV - 1.5 keV) the  ROSAT maps revealed numerous bright regions corresponding to hot gas within more distant cavities. One particularly interesting example is the so-called North Polar Spur (NPS), a prominent feature in the  X-ray sky, with conspicuous counterparts in radio continuum maps  (\cite{1964MNRAS.127..273H}). Despite its unique angular size, brightness,  and shape, its location is still an open question. 
It is thought to be the X-ray counterpart of a nearby cavity blown by stellar winds from the Scorpio-Centaurus OB association, which lies at a distance of $\sim$170 pc (\cite{1995A&A...294L..25E}). This is in agreement with a study of the shells and global pattern of the polarized radio emission by \cite{woll07}, who modeled the emission as due to two interacting expanding shells surrounding bubbles. The distance to the Loop I/NPS central part is found to be closer than 100 pc, and in both scenarios the source region is supposed to be wide enough to also explain the X-ray and the radio continuum enhancements  seen in the south, as a continuation of the Northern features. 
In complete opposition,  \cite{2000ApJ...540..224S} suggested that the NPS has a Galactic center origin, based on models of bipolar hypershells due to strong starburst episodes at the Galactic center, that  can well reproduce the radio NPS and related spurs. The debate has been recently reactivated, since new studies of NPS spectra have contradictorily either reinforced or questioned the local interpretation. \cite{willi03} have modeled XMM-Newton spectra towards three different directions and found evidence for a 280 pc wide emitting cavity centered 210 pc from the Sun towards $(l,b)=(352^\circ,+10^\circ)$, based on their determinations of the foreground and background IS columns. On the other hand, \cite{miller08} measured a surprisingly large abundance of Nitrogen in Suzaku spectra of the NPS, and interpreted it as due to AGB star enrichment, in contradiction with the Sco-Cen association origin. Finally, huge Galactic bubbles blown from the Galactic center in the two hemispheres were found in gamma rays (the so-called Fermi bubbles) and in microwave (WMAP Haze Bubbles). The bubbles are surrounded at low latitudes by gamma-ray and microwave arcs, and the NPS looks similar to the more external arc, suggesting an association. Uncertainties on the distance to the source region of the NPS remain an obstacle to the solution and end of debate.

3D maps of the ISM providing distances, sizes, and morphologies of the cavities together with the locations of dense clouds and new-born stars,  should further the analysis of the X-ray data. One of the ways to obtain a 3D distribution of  the ISM is to gather a large set of individual interstellar absorption measurements or extinction measurements toward target stars located at known and widely distributed distances, and to invert those line-of-sight (LOS) data using a  tomographic method. 
Based on the general solution for non linear inverse problems developed in the  pioneering work  of \cite{tarantola}, \cite{vergely01} developed its application to the nearby ISM and to different kinds of IS tracers, namely NaI, CaII, and extinction measurements (\cite{lall03,vergely10,welsh10}). Those studies, that used Hipparcos distances, provided the first computed 3D maps of the nearby ISM. Due to the limited number of targets those maps have a very low resolution and are restricted to the first 200-250 parsecs from the Sun. 
Recently, \cite{lallement13} compiled a larger dataset of $\sim$23,000 extinction measurements from several catalogs, and using both Hipparcos and photometric distances updated the 3D distribution.  The new 3D map has a wider coverage, showing structures as distant as 1.5 kpc in some areas.
Inverting this dataset paradoxically provides a better mapping of the nearby cavities than nearby clouds,  because the photometric catalogs are strongly biased to small extinctions. A number of cavities of various sizes, including the Local Bubble (LB) are now mapped and found to be connected to each other. Since most cavities are supposedly filled with hot, X-ray emitting gas, such 3D distributions are appropriate for further comparisons with diffuse X-ray background data, and this work is an attempt to illustrate examples of potential studies based on the combination of 3D maps and diffuse X-ray  emission. 

We address in particular the two questions previously mentioned: (i) we compare the shapes and sizes of the nearby cavities  with the soft X-ray data, in a search for properties of the nearby hot gas; (ii) we search a potential source region for the NPS emission in the new 3D maps.

In section 2, we present the inversion results in the Galactic plane, a  model of the soft X-ray emission based on this inverted distribution  and its comparison with the ROSAT 0.25 keV  emission. Uncertainties introduced by the correction from the heliospheric emission are discussed. 
In section 3, we search for IS structures in the 3D distribution that could correspond to X-ray bright regions and to the NPS source region. We use ROSAT survey images, Newton X-ray Multi-Mirror Mission (XMM-Newton) X-ray spectra and the 3D maps and discuss the possible location and origin of the NPS. We discuss our conclusions in section 4. 


\section{Local cavities and inferences about nearby hot gas}

\begin{figure}[htbp]
\begin{center}
\includegraphics[width=\linewidth]{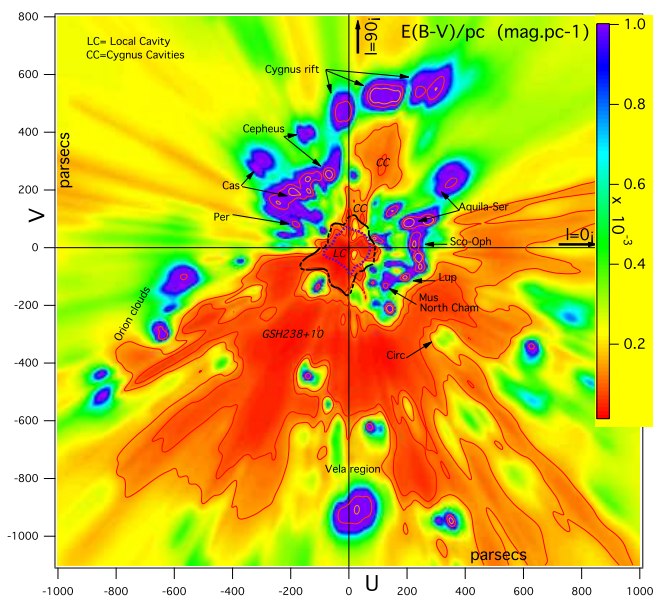}
\caption{Differential color excess in the Galactic plane, derived by inversion of line-of-sight data (map taken from \cite{lallement13}). The Sun is at (0,0) and the Galactic center direction is to the right. Cavities (in red) are potential soft X-ray background sources if they are filled with hot gas. A polar plot of the unabsorbed (foreground) 0.25 keV diffuse background derived from shadowing (\cite{1998ApJS..117..233S}) is shown superimposed in black (thick line), centered on the Sun. The linear scaling of the X-ray surface brightness, i.e. the ratio between surface brightnesses and parsecs is chosen to be consistent with the physical extent of the LB. Also shown is the shape of  the estimated average heliospheric contribution to the signal (dashed blue line). This contribution has been arbitrarily scaled to make it more visible and the heliospheric signal should not be directly compared here with the measured background.}
\label{galplane}
\end{center}
\end{figure}

The Local Bubble (LB) or Local Cavity is defined as the irregularly shaped, $\sim$ 50-150 pc wide volume of very low density gas surrounding the Sun, mostly filled by gas at $\sim$1 million Kelvin (MK). 
In most directions, it is surrounded by walls of denser neutral ISM, while in other directions, it is extended and connected through \textit{tunnels} to nearby cavities (see Fig. \ref{galplane}). 
Such a \textit{foamy}  structure is expected to be shaped by  the permanent ISM recycling under the action of stellar winds and supernova (SN) explosions that inflate hot MK bubbles within dense clouds. Hydrodynamical models of the multi-phase ISM evolution under the action of stellar winds and supernova remnants (SNRs, e.g. \cite{avillez09}) demonstrate how high density and temperature contrasts are maintained as shells are cooling and condense over the following thousand to million years, giving birth to new stars. Clouds engulfed by expanding hot gas evaporate. However, depending on their sizes and densities they may survive up to Myrs in the hot cavities. This is the case for the LB: as a matter of fact, beside hot gas, cooler clouds ($\sim$ 5-10 pc wide) are also present inside the bubble, including a group of very tenuous clouds within 10-20 parsecs from the Sun, the \textit{Local Fluff}. 

Hot gas ($T\sim10^6$ K) emits soft X-rays and produces a background that has been well observed during the ROSAT survey (\cite{1995ApJ...454..643S, snowden97}). 
The $10^6$ K gas is primarily traced in the 0.25 keV band, while hotter areas appear as distinct enhancements in the 0.75 keV maps, whose sources are supernova remnants, young stellar bubbles or super-bubbles (\cite{1998ApJ...493..715S}).  In the softer bands a fraction of the X-ray emission is generated in the Galactic halo, and there is also an extragalactic contribution. 

With the assumption that the Local Bubble and other nearby cavities in the ISM are filled with hot gas, then in principle knowledge of the 3D geometry of the ISM may be used to synthesize soft X-ray background intensity distributions (maps), as well as spectral information to compare with observations. This can be done by computing the X-ray emission from all empty (low density) regions, and absorption by any matter between the emitting region and the Sun. 
Provided maps and data are precise enough, it should be possible to derive the pressure of the hot gas as a function of direction and partly of distance. This is what we illustrate here, in a simplified manner, using only the X-ray surface brightness. This work will be updated and improved as higher-resolution 3D maps become available.

Figure \ref{galplane} shows the large-scale structure of the local ISM as it comes from the inversion of stellar light reddening  measurements. About 23,000 color excess measurements based on the Str\"omgren, Geneva and Geneva-Copenhagen photometric systems  have been assembled for target stars located at distances distributed from 5 to 2,000 pc. 
The photometric catalogs, the merging of data and the application to this dataset of the robust inversion method devised by \cite{tarantola} and developed by \cite{vergely01} and \cite{vergely10}  are described by \cite{lallement13}.
This inversion has produced a 3D distribution of IS clouds in a more extended part of the solar neighborhood compared to previous inversion maps. E.g., clouds in the third quadrant are mapped as far as 1.2 kpc.
On the other hand, this dataset is still limited and precludes the production of high resolution maps. The correlation length in the inversion is assumed to be of the order of 15 pc, in other words all structures are assumed to be at least 15 pc wide. 
As a consequence of this smoothing and of the very conservative choice of inversion parameters (made in order to avoid artefacts), the maps do not succeed in revealing the detailed structure of the local fluff in the Solar vicinity, whose reddening is very low due to both a small gas density and a small dust to gas ratio. 

Along with the mapping of the dust clouds, quite paradoxically those datasets are especially appropriate to reveal voids (i.e. cavities), essentially due to a bias that favors weakly reddened stars in the database. The maps indeed show very well the contours of the LB, seen as a void around the Sun, and a number of nearby empty regions (\cite{lallement13}). 
Among the latter, the most conspicuous mapped void by far is a kpc wide cavity in the third quadrant that has been identified as the counterpart of the GSH238+08+10 supershell seen in radio (\cite{heiles98}). It is a continuation of the well known $\beta$CMa empty, extended region (\cite{1985ApJ...296..593G}). It is of particular interest here as a potential source of strong diffuse X-ray emission.

 In the following we make use of two different quantities derived from ROSAT data: (i) the initial data measured in the R1-R2 channels that correspond to the 0.25 keV range, after correction for the LTE's, and (ii) the unabsorbed emission in the same energy range, a quantity computed by \cite{1998ApJ...493..715S} obtained after subtraction of any source that can be shown to be non negligibly absorbed by foreground IS clouds. This residual emission from hot gas in the solar neighborhood that is not hidden by dense clouds has been determined by computing the negative correlation between the R1-R2 corrected background and the 100 $\mu$m intensity measured by the Infrared Astronomical Satellite (IRAS), corrected using data from the Diffuse Infrared Background Experiment (DIRBE) (\cite{SFD98}). 

An important question is the fraction of heliospheric emission that has not been removed to obtain the true signal, based on time variability. It is a difficult task to estimate it precisely, and this will be addressed in future works using the most recent heliospheric models. Since our purpose here is to illustrate the potentialities of the 3D maps,  we will use the published estimates of the solar wind contribution to the cleaned signal (\cite{lall04}), and allow for a large uncertainty, that will also cover the small geocoronal zero level remaining contribution. 


\subsection{Galactic Plane: morphological comparisons}

\begin{figure*}[t!]
\begin{center}
\includegraphics[width=0.8\linewidth]{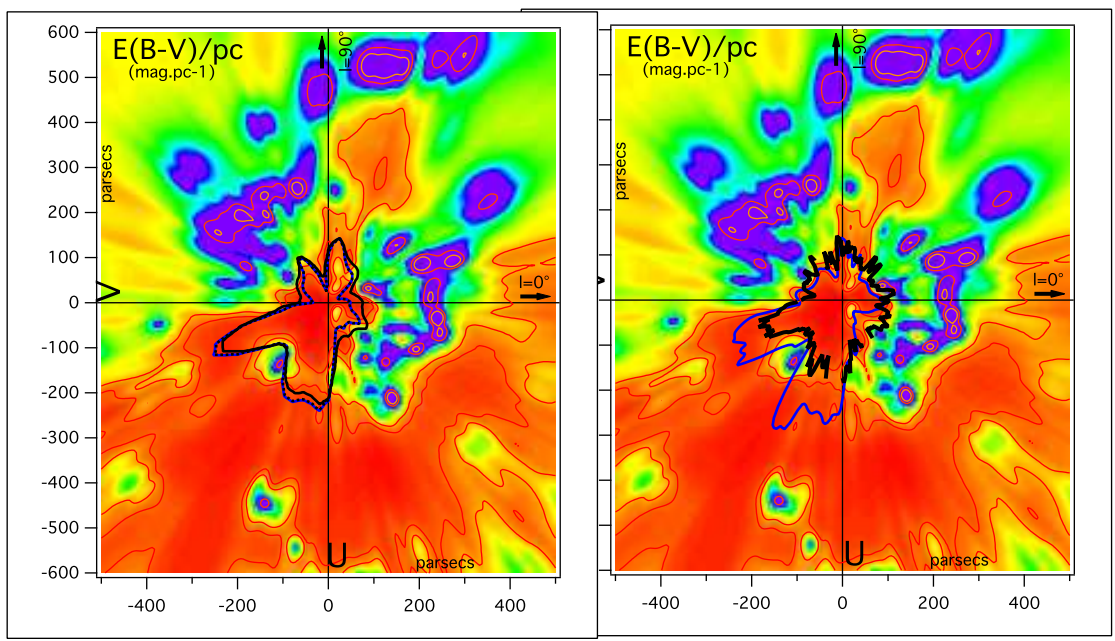}
\caption{Same as Fig. \ref{galplane} in a restricted area around the Sun. The color scale is identical to the one in Fig \ref{galplane}. Left: A polar plot centered on the Sun shows the  unabsorbed 0.25 keV diffuse background after removal of a heliospheric contribution. The black and blue curves correspond to different assumptions on the actual level of the heliospheric foreground (see text). Right: Polar plots representing the total 0.25 keV background (absorbed + unabsorbed), after removal of the larger heliospheric contribution (see black line), and the result of a simplistic radiative transfer model assuming all cavities are filled with hot, homogenous gas (blue line). The scaling for the polar plot curves is done in the same manner as in Fig. 1.}
\label{galplane2}
\end{center}
\end{figure*}

As mentioned above, we are using recent 3D ISM maps resulting from the inversion of color-excess measurements. The inversion provides the distribution of the differential opacity in a cube, and we generally use slices within this cube to study the IS structures in specific planes.
Figure \ref{galplane} shows a horizontal slice of the 3D inverted cube that contains the Sun. The differential extinction is color-coded, in red for low volume opacity and purple for high volume opacity. Despite the small offset of the Sun from the Plane, we consider this slice as corresponding to the Galactic plane.  The map shows distinctly the $\sim$ 50-150 pc wide empty volume around the Sun that corresponds to the LB. The huge cavity in the third quadrant is the largest cavity that is directly connected with the LB. We note that the direction of this conspicuous cavity (also, as noted above, of the closer $\beta$CMa tunnel) corresponds to the direction of the helium ionization gradient found by  \cite{wolff99}, suggesting that the same particular event might be responsible for the formation of the big cavity and also influenced the ISM ionization close to the Sun. The cavity is partially hidden behind a dense IS cloud located at $\sim$ 200 pc towards $l = 230^\circ$. We also see an extended cavity (presumably a hot gas region) at $l \sim 60^\circ$,  and smaller, less well-defined extensions of the LB towards $l \sim 285^\circ$ and $l \sim 340^\circ$.

Superimposed on the opacity map is the ROSAT unabsorbed 0.25 keV surface brightness derived by \cite{1998ApJ...493..715S}, shown in Sun-centered polar coordinates.  Although the curve represents surface brightness values and not distances, such a representation reveals the correspondence between X-ray bright regions and cavity contours. It is immediately clear that the two brightest areas correspond to the regions in the third quadrant that are not hidden by the $\sim$200 pc cloud, suggesting that a fraction of the signal comes from the wide GSH238+10 (or CMa super-bubble) cavity, and that towards the cloud this signal is attenuated. Other enhancements correspond to directions where the LB first boundary is quite distant ($l \sim +80^\circ, +120^\circ, +330^\circ$), strongly suggesting that part of the emission is associated to those elongated parts of the LB. Those coincidences are quite encouraging, showing that the mapped cavities and the soft X-ray emission are correlated.
  
Figure \ref{galplane} also displays the shape of the averaged estimated heliospheric contribution to the signal, computed using the ROSAT observing conditions (values taken from \cite{lall04}). Because this signal is not far from being isotropic in the galactic plane, removing it from the total background results in an amplification of the enhancements previously mentioned. This can be seen in Fig \ref{galplane2} which shows the residual signal after removal of the same heliospheric pattern, for two different assumptions about the absolute level of this contamination (see \cite{lall04}). Considering those two cases also allows to figure out how the subtraction of a small zero level of geocoronal contribution would impact on the hot gas emission pattern. In both cases, by using an appropriate X-ray brightness scaling for the figure, we can see that the corrected background shape agrees quite well with the LB contours, following the empty regions derived from the inversion results.  On the other hand, this also shows the sensitivity to the heliospheric contribution and the importance of taking this heliospheric counterpart into account as accurately as possible.  


\subsection{Galactic Plane: a simplified radiative transfer model}

Based on our 3D maps, we have estimated the X-ray intensity ($I$) observed at the Sun location, assuming all cavities are X-ray emitters. We integrate the emission from the Sun to a given distance $D$. Here we use D = 1500 pc, but computations have shown that above D=300 pc and for most directions the average brightness measured at the Sun does do not change significantly, since the major part of the emission generated at large distance is re-absorbed by the closer matter. For this first attempt we assume that all cavities (or low density) regions have the same X-ray volume emissivity, proportional to a coefficient ($\epsilon$).
As a criterion to distinguish between a cavity (X-ray emitter) and a cloud, we assume that there is X-ray emission ($\delta=1$) everywhere the differential extinction is lower than a chosen threshold, and that there is no emission 
everywhere the density is higher than 
this threshold value. We have selected here the threshold $\rho_0 < 0.00015$ mag.pc$^{-1}$, that corresponds well to the marked transition between the dense areas and the cavities (this corresponds to orange color in Figs. \ref{galplane} or \ref{galplane2}). Finally, we also assume that at any location  the opacity to X-rays is proportional to the dust opacity in the optical (which implies that the gas in the mapped cavities is also absorbing, however since the dust opacity is very small this is a negligible effect). 

Our assumption of a two-phases medium, one emitting X-rays and one absorbing them, corresponds to an over-simplified situation. Indeed, hydrodynamical models of the ISM physical and dynamical evolution under the action of stellar winds and supernovae show that these processes lead to extremely complex structures, and, especially, show that a large fraction of gas that is out of collisional equilibrium and can not be part of the two categories (see a review of the models by \cite{avillez13}).  Regions that have been recently  shock-heated are filled by under-ionized gas that does not emit in X-rays despite their high electron temperature, and conversely gas rapidly expanding may be cooling so fast that ions remain in charge states much above the equilibrium levels and continue to contribute to the X-ray background \citep{avillez09,avillez12}. The first case is more important here, since it implies that we may map cavities that are made of emitting and non-emitting areas. In order to take into account this possibility, we introduce a filling factor f$_{x}$ for the X-ray emitting gas, in such a way the fraction (1-f$_{x}$) may represents the low density, hot under-ionized gas. It may also represent low dust, warm ionized gas, if any. Unfortunately, at this stage of the modeling we can not address the complex physics of the multi-phase IS medium.

Integrating from large distance towards the Sun, the X-ray brightness obeys the simple differential equation:
\begin{equation}
\frac{dI}{ds}=  \delta  f_{x}   \epsilon - I   k  \rho
\end{equation}
where the first term accounts for the emission generated along $ds$, IF and only IF there is hot gas at distance $s$, i.e. $\delta=1$ if  $\rho \leq \rho_0$ threshold and $\delta=0$  if  $\rho \ge \rho_0$, and the second term accounts for the absorption along the distance $ds$, assumed to be proportional to the differential color excess $\rho$.  $k$ is a coefficient converting the differential color excess E(B-V) into the 0.25 keV differential opacity.

The coefficient $k$ is estimated in the following way: in the visible $E(B-V)_{dust}=1$ corresponds to $N(H) \sim 5 \times 10^{21}$ cm$^{-2}$, while, at 0.25 keV,  $\tau_{0.25 keV}=1$ corresponds to  $\sim 1 \times 10^{20}$ cm$^{-2}$. We thus simply assume here that the coefficient $k$ is of the order of 50. 

The integral is computed by discretizing the line-of-sight 
into 1 pc bins. The differential opacity $\rho$ at the center of each bin is obtained by interpolation in the 3D cube. The integral  is computed inwards, from the distance $D$ to the Sun location. 

The resulting brightness is displayed as a polar plot and superimposed on the galactic plane map in Fig \ref{galplane2} (right). It can be seen that the model is closely resembling in shape the unabsorbed emission computed by \cite{1998ApJS..117..233S} displayed in the left panel, which constitutes an interesting validation of the mapping computation. However, because the simple radiative transfer model involves emission from all regions, and includes partially absorbed signals, the data-model comparison should be made with the initial ROSAT data. We thus extracted the 0.25 keV surface brightness  in the galactic plane from ROSAT high resolution maps, using the $b = +/-1.5^\circ$ slice in the data.  We exclude data in the direction of Vela (at $l \sim 260^\circ$) due to the strong SNR contribution to the observed flux. We show in polar coordinates the resulting signal,  after removal of the same heliospheric contribution as the one used above. Again there are strong similarities between the angular shapes of the model and the data, with however an interesting, significant overestimation of the model towards the CMa cavity.  We will return to that point in the discussion below. An important result of this comparison is the good correspondency between the elongated parts of the Local
Cavity and soft X-ray maxima, i.e. towards the longitudes $l \sim $ 345, 80 and 125$^{\circ}$. This demonstrates that there is X-ray emitting gas filling those nearby regions that are considered as part of the Local Bubble, and strongly suggests that hot gas is filling the entire Local Cavity, except for the small clouds known to be embedded, e.g.  the \textit{Local Fluff}.

A quantitative comparison between the model and the data requires some assumptions about both the emitting gas properties and the instrument sensitivity. Here we make use of the estimates of Snowden et al. (1998), namely that a 1 MK gas with solar metallicity produces 1 count s$^{-1}$ arcmin$^{-2}$ per $7.07$  cm$^{-6}$ pc emission measure, for a spectral dependence corresponding to the thermal equilibrium model of \cite{1977ApJS...35..419R}.
Using this relationship and masking the Vela region at $l=-100^\circ$, the average ROSAT X-ray brightness in the Plane  is 0.0003 counts s$^{-1}$ arcmin$^{-2}$, or 0.002 cm$^{-6}$ pc. On the model side, the average calculated intensity from the simplified radiative transfer computation is $I =112 \times f_{x}$  for $\epsilon=1$ and using parsecs as distance units, and, if we exclude the third quadrant, $I =78 \times f_{x}$. 
Within the assumption of a homogeneous hot gas, this computed quantity is equivalent  to an emission measure if $\epsilon=n_{e}^{2}$, $n_{e}$ being the electron density. 
Equating data and model, we have: 
$EM_{x-ray} = n_{e}^{2} \times I \times f_{x} $ which implies  then $n_e = \sqrt{0.002/112}= 4.2 \times 10^{-3}$  $\times f_{x}^{-\frac{1}{2}}$ cm$^{-3}$ when averaging over all longitudes, and $n_e = \sqrt{0.002/78}= 5.1 \times 10^{-3}$  $\times f_{x}^{-\frac{1}{2}}$ cm$^{-3}$ when excluding the third quadrant.
The corresponding average pressure for a 1 MK gas is $P= 2 \times n_e \times T = 8, 400 f_{x}^{-\frac{1}{2}}$ K cm$^{-3}$ when considering the four quadrants or $10, 200 f_{x}^{-\frac{1}{2}}$  K cm$^{-3}$ when excluding quadrant III. Assuming a filling factor of 1 and excluding quadrant III, i.e. restricting the adjustment to the Local Cavity itself, the derived average pressure P= $10, 200$  K cm$^{-3}$ is reasonably close to what has been inferred previously (15,000 K cm$^{-3}$, \cite{1998ApJ...493..715S}), showing here that this new method based on the extinction maps provides reasonable results, despite the numerous assumptions made here such as a canonical dust/gas ratio and the hot gas temperature homogeneity. Further work however is needed to improve the knowledge of the SWCX contribution, since underestimating this contribution would result in a lower hot gas pressure.

\begin{figure*}[t!]
\begin{center}
\includegraphics[width=0.6\linewidth]{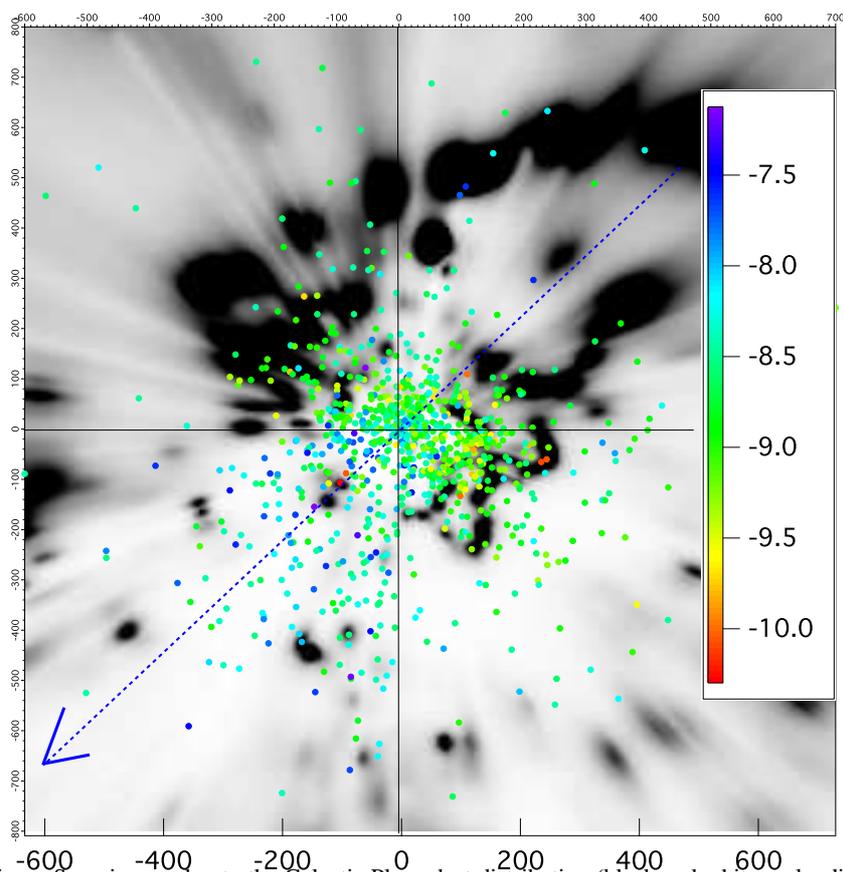}
\caption{Dust-CaII comparison. Superimposed onto the Galactic Plane dust distribution (black and white scale, distribution identical to the one in Fig \ref{galplane}) are the projections of those stars located within 100 pc from the Plane and for which CaII columns have been measured in absorption (colored dots). Units are in parsecs from the Sun (at center). The dot color represents the ratio between the CaII column measured towards the star and the hydrogen column that corresponds to its color excess. The color excess is computed by integrating the 3D differential color excess from the Sun to the star location. The hydrogen column is assumed to be 5 10$^{21}$ x E(B-V) in cm$^{-2}$ units. The color code refers to the logarithm of this ratio, in cm$^{-2}$.mag$^{-1}$. Stars in the large cavity in the 3rd quadrant have a particularly high CaII to dust (or equivalent H) ratio. The arrow marks the ionization gradient direction found by \cite{wolff99}.}
\label{caIItodust}
\end{center}
\end{figure*}

We have seen that under the assumption of hot gas homogeneity, and if the filling factor is assumed to be of the order of unity in all directions, then the model overestimates significantly the emission in the third quadrant. This discrepancy is not due to an absence of emitting gas beyond 100-150 pc, as demonstrated by the shadowing effect of the $l \sim 230^\circ$, $d \sim 200$ pc cloud discussed above. Instead, this discrepancy is very likely explained by different physical properties in this part of the sky, namely the existence of large quantities of warm gas, or equivalently of a filling factor significantly smaller than one. \cite{snowden98} already noted that there is a large quantity of warm, ionized gas in this area and that this gas may occupy a significant fraction of space. In our simple dichotomy of cavities and dense clouds, we overlooked the role of such ionized, low density and low dust regions. In order to investigate this potential role, we have used in conjunction both our database of ionized calcium absorption data (data recently compiled by \cite{welsh10}) and the 3D dust maps: for all stars for which interstellar ionized calcium has been measured, we integrated from the Sun to the target star through the 3D differential dust opacity distribution, to obtain the total line-of-sight opacity. We then computed, for each target star, the ratio between this integrated dust opacity and the CaII column. Results are displayed in Fig. \ref{caIItodust} for all stars that are located within 100 parsecs from the Plane. The figure clearly shows that the third quadrant is strikingly depleted in dust, or enriched in CaII, or both simultaneously. As a matter of fact, the CaII to extinction ratio is larger by more than an order of magnitude than the corresponding ratio in the three other quadrants. Indeed, it is well known that in shocked and heated gas dust grains are evaporated and calcium is released in the gas phase, two effects that both contribute to increase the CaII to dust ratio. It is thus clear that our assumption that every region with very low opacity is filled by hot gas is equivalent to neglecting the volume occupied by ionized and heated gas (but not hot enough to emit X-rays) and that may be the source of the observed deviation in the third quadrant. For stars located in the third quadrant at about $\sim$300 pc or more, CaII columns reach 10$^{12}$ to 10$^{13}$ cm$^{-2}$. Approximate conversions to distances by means of the average values found for the local clouds (see e.g. \cite{redfield00}) results in large distances, between 100 and 1000 parsecs, confirming that a large fraction of the line-of-sight may be filled by warm ionized gas and not hot gas, and that $f_{x}\leq1$ in the third quadrant large cavity.
Moreover, our assumption of a homogeneous dust to gas ratio also results in underestimating the absorption by the warm dust-depleted ionized gas, which may also contribute to the discrepancies. Such results illustrate the need for additional various absorption data in the course of using 3D dust maps and diffuse X-ray background.

The fourth quadrant is also mapped as largely devoid of dust. However, at variance with the third one, there are very opaque regions at about 100 to 300 pc, which introduces a bias in the sense that very few targets are part of the database that are located beyond those opaque clouds. Not much can be said about this quadrant at large distance, except that according to the available targets the CaII to dust ratio is lower than in the big cavity of the third quadrant, suggesting that the event that has evaporated the dust and ionized the gas in the third quadrant has not operated in the same manner in the fourth quadrant. 

\subsection{Meridian plane}

\begin{figure}[htbp]
\begin{center}
\includegraphics[width=\linewidth]{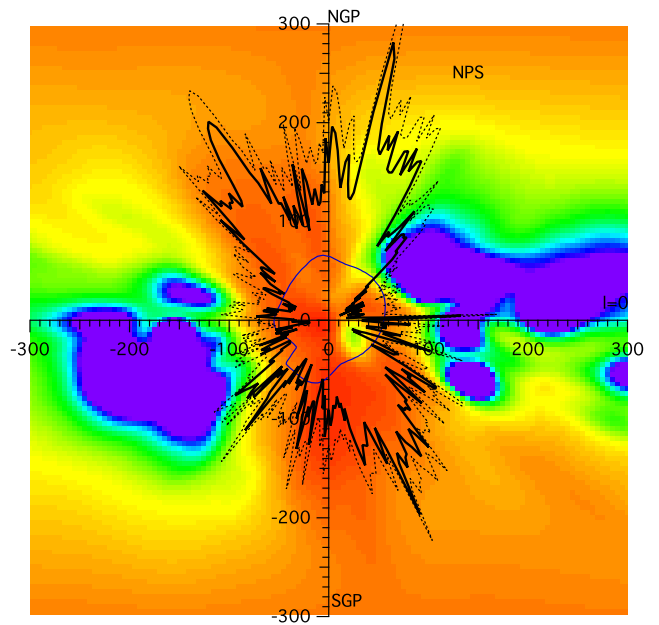}
\caption{Inversion result (differential opacity) in the meridian plane. The Sun is at (0,0). The color scale is identical to the one in Fig \ref{galplane}.  The galactic center is to the right and the North galactic Pole (NGP) to the top. Superimposed are a polar plot of  the 0.25 keV background (dashed black line), and a polar plot of the same emission after subtraction of the heliospheric counterpart (thick black line). The angular dependence of the heliospheric contribution is shown in blue. The jump in the region around $l,b=(180^\circ,-40^\circ)$ is due to the ROSAT measurement geometry (see text). The scaling for the polar plot curves is done in the same manner as in Fig. 1.  The spiked nature of the polar plot is due to the statistical variation of the X-ray data.}
\label{meridplane}
\end{center}
\end{figure}

Out of the Galactic plane, the comparison between the 3D maps and the soft X-ray background  is much more difficult. As a matter of fact the above study benefited from the following favorable conditions: (i) the 0.25 keV emission in the Plane is solely due to nearby hot gas as the nearest HI clouds are optically thick, and (ii) the maps have the greatest accuracy there, due to the large number of target stars used for the inversion.  At variance with the Plane, at high latitudes condition (i) is not fulfilled, because there is additional emission from the halo and extragalactic sources, nor condition (ii), because 3D maps are much more imprecise and limited in distance due to the poorer target coverage. For those reasons, and because the lack of precision is maximum in the rotation plane (plane that is perpendicular to meridian plane to Galactic center, or slice in $l$=90-270$^\circ$), we will restrict out-of-Plane studies to a brief comparison between the dust distribution and the soft X-ray background in the meridian plane. Figure \ref{meridplane} shows the meridian plane dust map resulting from the color excess data inversion, with the 0.25 keV emission (before and after subtraction of the heliospheric counterpart, also shown)  superimposed in polar plots. 

The emission has two well defined maxima along a  $\sim70 ^\circ$ inclined axis that corresponds to the two \textit{chimneys} to the halo, an axis that is known for being roughly perpendicular to the Gould belt plane. The plane itself is traced by the main masses of IS matter above and below the Plane towards $l=0^\circ$ and $l=180^\circ$ respectively. Such qualitative results were already obtained based on previous neutral sodium and extinction maps (\cite{lall03, vergely10}), and are simply confirmed using a larger number of targets. In those \textit{chimney} directions the signal must be decomposed into the extragalactic, halo, LB, and heliospheric contributions. This requires a careful analysis of the latter contamination, which will be the subject of future work.

On the other hand, the simple comparison between the the improved distribution and the X-ray pattern interestingly shows that the most conspicuous emission at $l=0^\circ$,  $b \sim 60-75^\circ$ that corresponds to the well known North Polar Spur/Loop1 enhancement does not correspond to a particular nearby ($\lesssim$ 300 pc) cavity. This lack of apparent nearby counterpart suggests a distant emission or a different emission mechanism for the NPS arch. We come back to this point in the next section that is entirely devoted to the NPS seen in the 0.75 keV range.

 
\section{The 0.75 keV diffuse background and 3D maps: search for the North Polar Spur (NPS) source region}

 \begin{figure*}[htbp]
\centering
\includegraphics[width=0.7\linewidth]{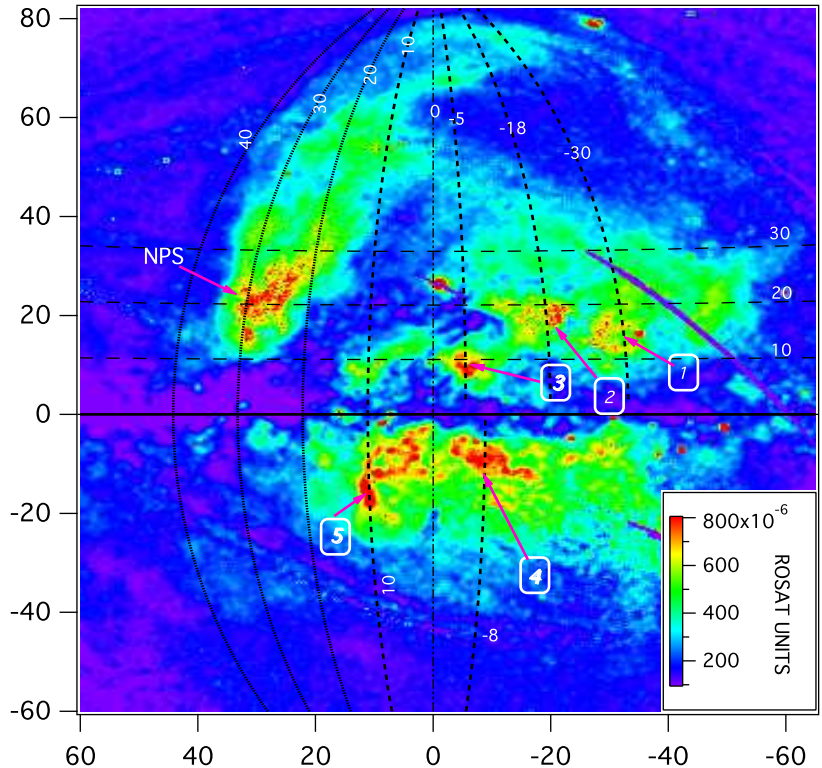}\\
\caption{ROSAT 0.75 keV surface brightness map around the Galactic Center. Five selected bright regions (red) are labeled and shown. The central parts of those regions correspond to the 5 line-of-sight drawn in the four meridional cuts in Fig \ref{fourslices} and the top-right of Fig \ref{findnps}, allowing to correlate visually the bright X-ray sources (hot ISM) and low density regions or tunnels appearing in the 3D maps.}
\label{corr}%
\end{figure*}


\subsection{0.75 keV bright regions in the first and fourth quadrants}

Figure \ref{corr} illustrates our first study of relationships between X-ray brightness enhancements areas and voids in the 3D ISM distribution. It shows a large fraction of the 0.75 keV ROSAT map centered on the GC direction. The 0.75 keV bright regions, in principle, correspond to relatively young ($\lesssim$ 1-2 Myrs) hot ISM bubbles blown by winds from newly born stars and SNRs, and they should correspond to voids in the 3D ISM distribution. Here, we attempt to use the inverted 3D distribution to study the correspondency between the bright 0.75 keV directions and the nearby cavities or cavity boundaries appearing in the maps. The study has some similarities with HI (or IR) vs X-ray anti-correlations or shadowing experiments. However here we use the local ISM distribution and not quantities integrated up to infinity, and our new approach is fully complementary. Our goal here is to infer in which way the NPS is similar (or different) from other bright features. 

We have selected the main bright regions in the Galactic center region besides the NPS which are wider than a few degrees, avoiding $b \lesssim 5^\circ$ region for which the emission is likely to be generated far outside our mapped volume (e.g., \cite{park97}). Numbered 1 to 5 in the following, their approximate centers are $l=330^\circ$ (-30$^\circ$), $l=342^\circ$ (-18$^\circ$), $l=355^\circ$ (-5$^\circ$) and $l=352^\circ$ (-8$^\circ$), and $l=10^\circ$. They are shown in Fig. \ref{corr}. 

\begin{figure*}[htbp]
\centering
\includegraphics[width=0.34\linewidth]{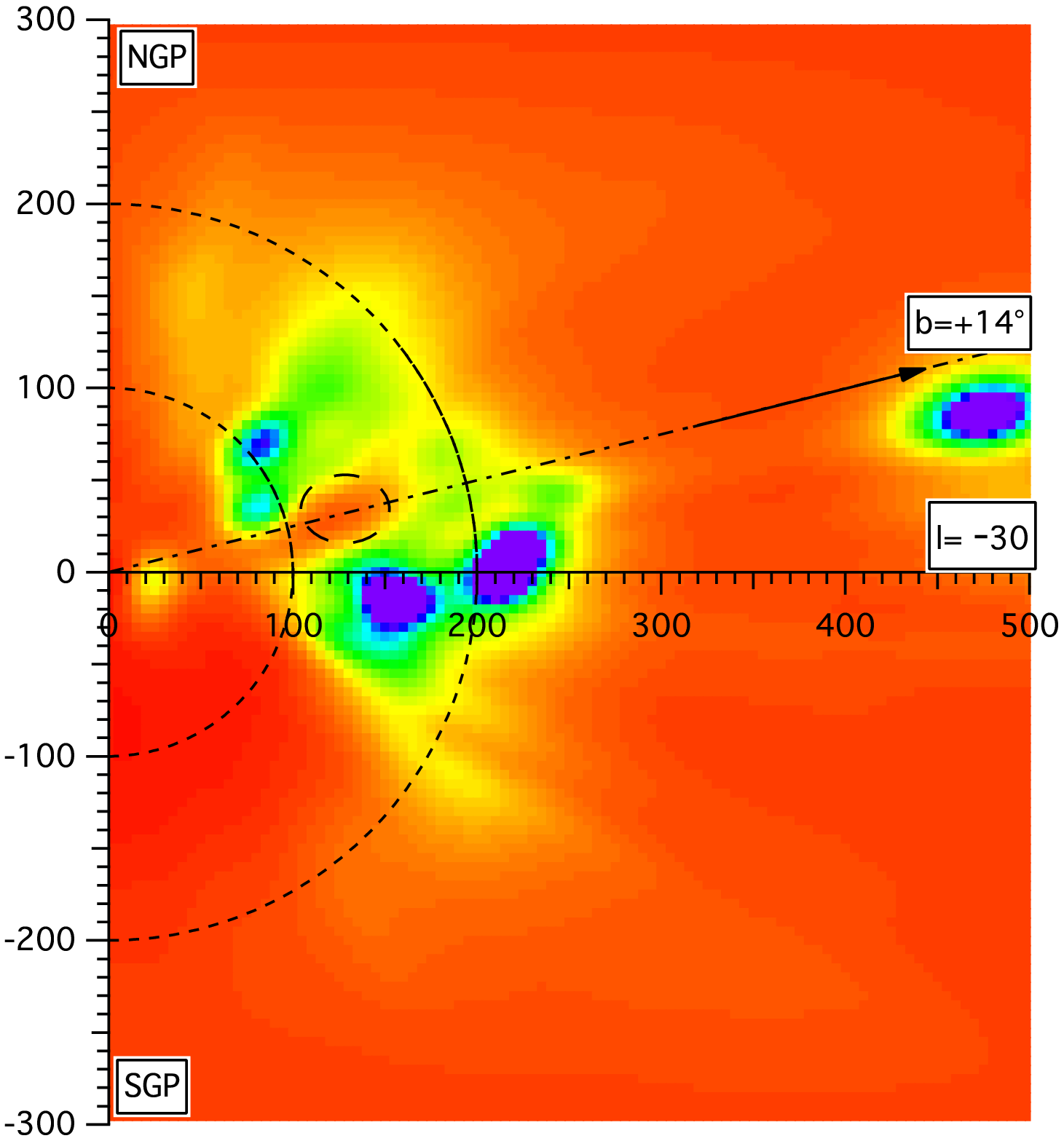}
\includegraphics[width=0.34\linewidth]{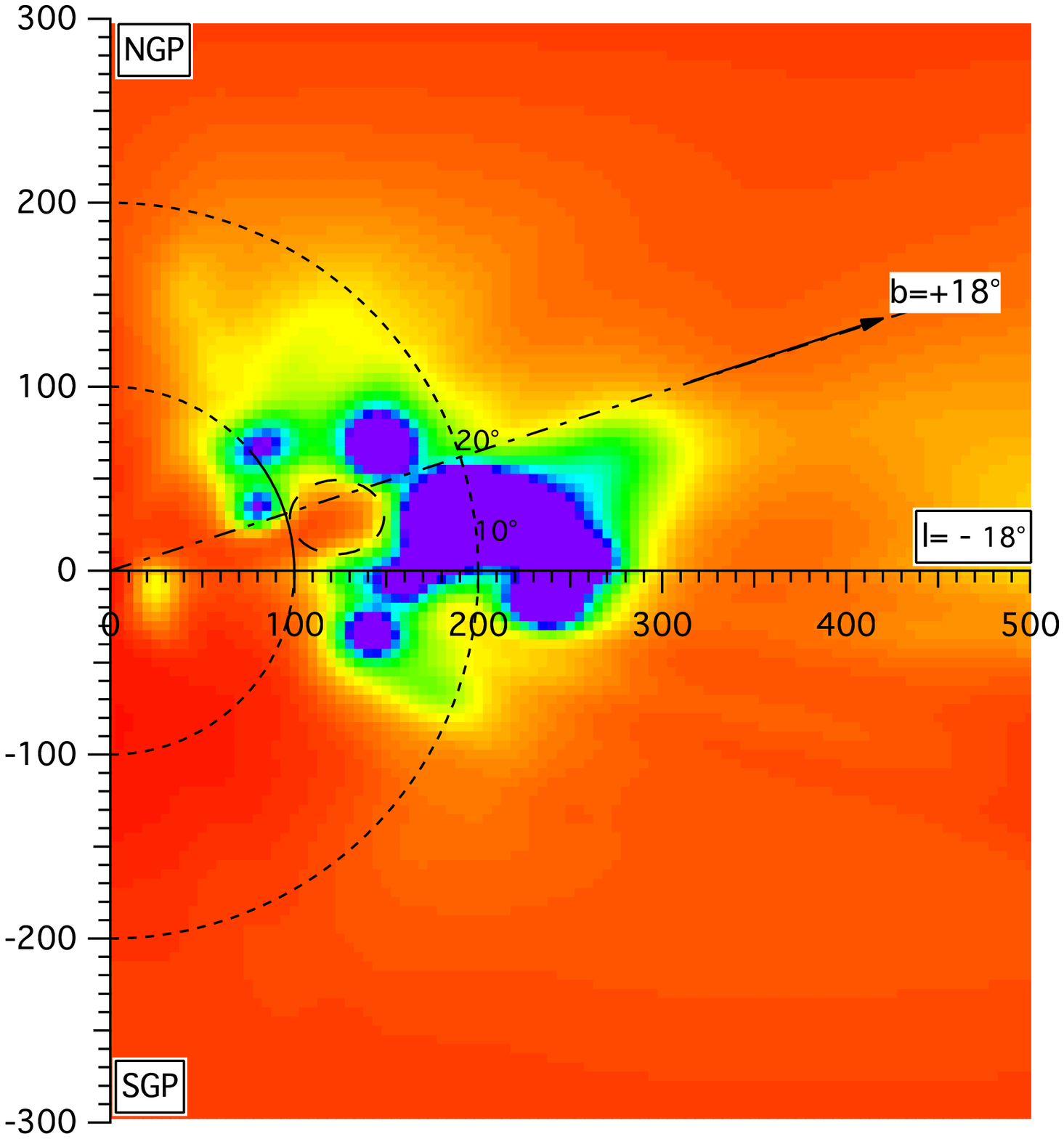}
\includegraphics[width=0.34\linewidth]{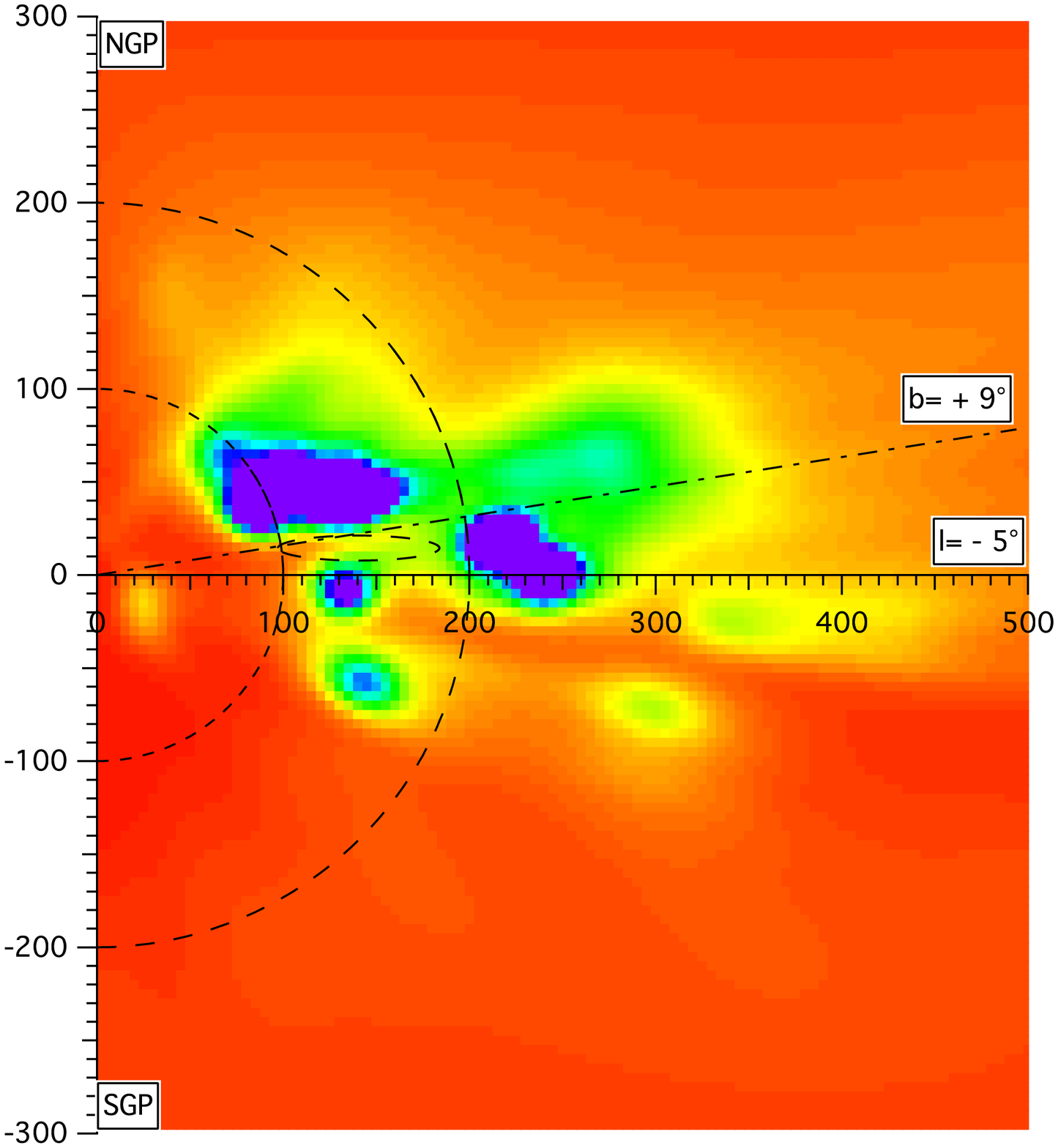}
\includegraphics[width=0.34\linewidth]{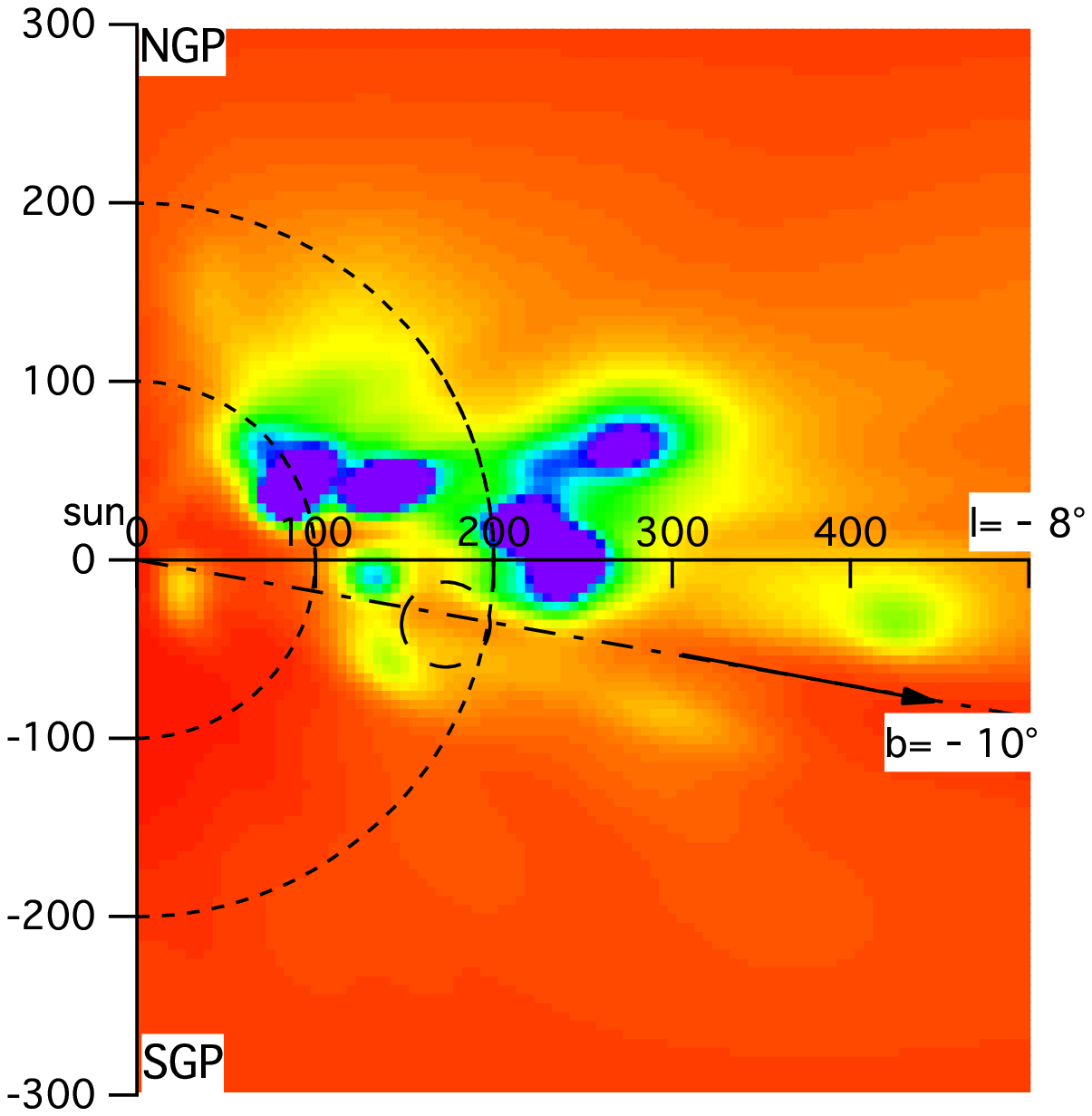}
\caption{Inversion results: vertical slices in the 3D distribution at $l=-30^\circ$ (case 1), $-18^\circ$ (case 2), $-5 ^\circ$ (case 3) and $-8^\circ$ (case 4). The first  vertical planar cut ($l=-30^\circ$) corresponds to part of the Figure 7 (middle) of \cite{lallement13}. The Sun is at (0,0) in those half-planes. The color scale is identical to the one in Fig \ref{galplane}. Sun-centered dashed black circles correspond to distances of 100 and 200 pc from the Sun respectively. The directions of the bright areas labeled in Fig. \ref{corr} are shown by a dot-dashed black line and an arrow. Dashed black ellipses mark cavities that may correspond to soft X-ray brightness enhancements of Fig. \ref{corr}.}
\label{fourslices}%
\end{figure*}

For the first four selected regions,  Fig. \ref{fourslices} displays the corresponding planar meridian cuts in the dust distribution, i.e. meridian planes that contain the central directions of those X-ray bright areas. The vertical slice corresponding to the fifth bright region at $l=10^\circ$, $b=-10^\circ$ to $-18^\circ$ is part of Fig \ref{findnps}. Arrows provide the latitude of the approximate centers of the bright regions deduced visually from the ROSAT map. 
For all five selected bright areas 
we find in the lower images (Fig. \ref{fourslices} and \ref{findnps}) a corresponding void region in the direction of the bright X-ray emission. The voids have the forms either of a tunnel linking to empty regions at larger distances, and thus corresponding to a potential source being located beyond the tunnel, or of a nearby cavity, as in the case of the $l=330^\circ$ enhancement. Such correspondences globally demonstrate a consistency between the local IS distribution (despite its poor resolution) and the soft X-ray maps. 
Note that interactive 3D images showing the densest dust structures can be seen at \\
\url{http://mygepi.obspm.fr/~rlallement/ism3d.html}    \\
and\\
\url{http://mygepi.obspm.fr/~rlallement/ism3dcrevace.html} \\
instead of using the drawn planar cuts.

\subsection{Search for a nearby cavity as a source region of the NPS}

The North Polar Spur (NPS) is a well-known conspicuous region of strong diffuse X-ray emission, which coincides in direction with the gigantic radio feature called Loop I.  Loop I was recognized at first as a Galactic giant radio continuum loop of $58^\circ \pm 4^\circ $ radius centered at $l=329^\circ \pm 1.5^\circ$ and $b=+17.5^\circ \pm 3^\circ$ \citep{1971A&A....14..252B}. The bright NPS is centered at $l \approx 30^\circ$ and extends north above $b \approx +10^\circ$. It has been very well mapped in the ROSAT observations (0.1-2.0 keV, \citep{1995ApJ...454..643S}). The NPS is believed to have a local origin, associated to a nearby super-bubble centered on the Sco-Cen OB associations ($\sim100$ pc),
the NPS and Loop I corresponding to the external parts and boundaries of this nearby super-bubble. Some kind of interaction might be at present between  the Local Bubble's wall in the Galactic center direction and the closest external region of this super-bubble. At higher latitude, faint filaments seen in HI and extending up to $85^\circ$ also coincide with the radio continuum shells.
At least the high latitude HI filaments in the fourth quadrant are very close to us. As  a matter of fact, using optical spectra, \cite{2012A&A...545A..21P} have set a distance of $98 \pm 6$ pc for those low velocity  shells.  This proximity is in good agreement with the Loop1-LB scenario. Other HI structures with higher radial velocities and in the same sky region are located beyond 200 pc. 



 \begin{figure*}[htbp]
\centering
\includegraphics[width=0.75\linewidth]{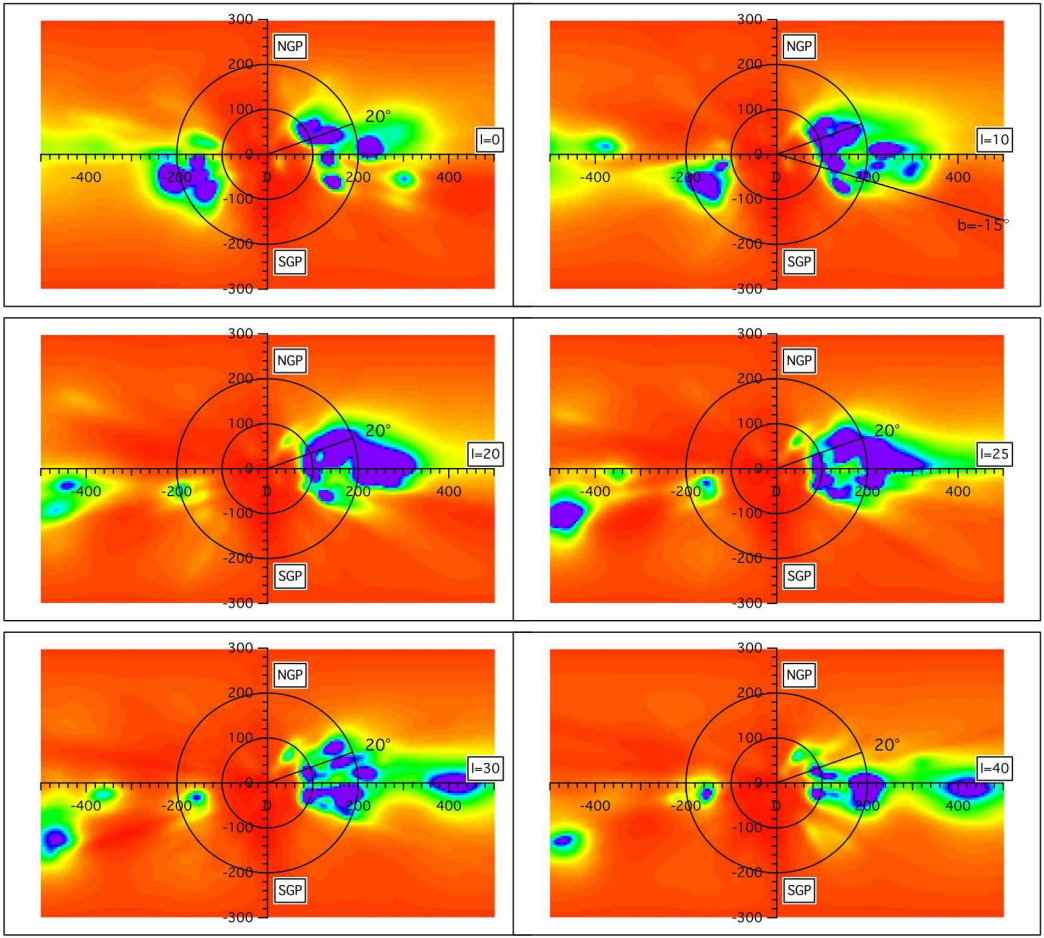}\\
\caption{Inversion result: vertical slices in the 3D dust distribution containing or around the NPS bright directions. The color scale is identical to the one in Fig \ref{galplane}.  The $l=+10^\circ$ slice contains the direction of the bright region number 5 of Fig \ref{corr}. Sun-centered black circles correspond to distances of 100 and 200 pc from the Sun respectively.The vertical planar cuts for $l=0-180^\circ$ and $l=30-210^\circ$  correspond to the Figure 4 (top and bottom) of \cite{lallement13}.}
\label{findnps}%
\end{figure*}

A contradictory, and indeed completely opposite view is also defended. According to it, the NPS is a fraction of a super-shell centered at the Galactic center  \citep{2003ApJ...582..246B,2000ApJ...540..224S}. More recently, the debate about the distance to the NPS source region has been reactivated due to the discovery of the Fermi \textit{bubbles} and associated gamma-ray and microwave arches associated to the bottom and envelops of the bubbles. As a matter of fact there is a similarity between the  shape of the NPS and arcs seen in gamma rays and microwave that seem to envelop the northern Fermi bubble. Hence, it is worthwhile to revisit the problem of the location of NPS.

\begin{figure*}[htbp]
\centering
\includegraphics[width=0.7\linewidth]{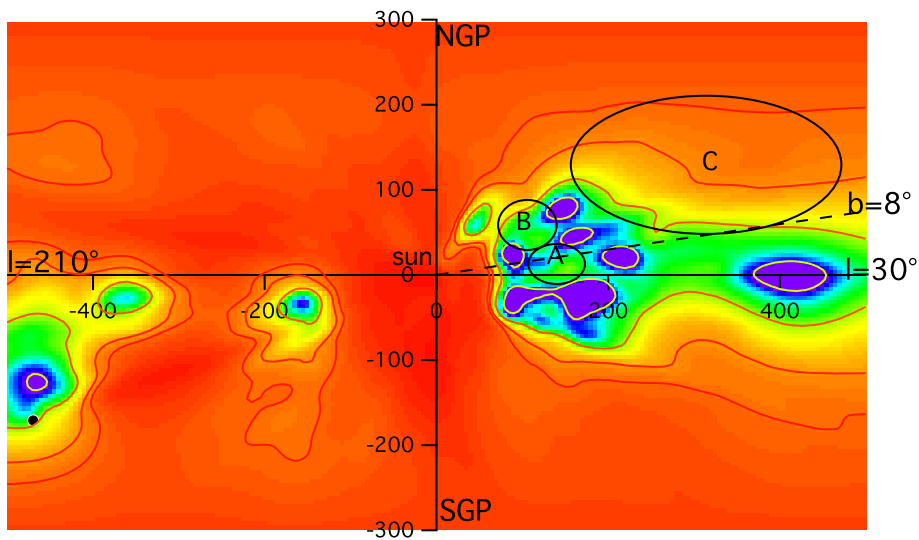}
\caption{Volume opacity within the $l=30-210^\circ$ vertical plane (identical to Fig. 4 (bottom) of \cite{lallement13}). The Sun is at (0,0). The color scale is identical to the one in Fig \ref{galplane}. The circles show potential locations of the NPS source region discussed in the text, assuming the NPS emission is generated within a cavity. The dashed line indicates the latitude b=+8$^\circ$, above which is detected the NPS.}
\label{npsvert30}%
\end{figure*}

One of the puzzling characteristics of the NPS is its abrupt disappearance below $b=+8^\circ$, seen in all ROSAT channels. Thus either the source region is extending below this latitude but is masked by a high column of gas at $b\lesssim+8^\circ$ (case 1), or the emission is generated only above $b=+8^\circ$ (case 2). We show in Fig. \ref{findnps} a series of vertical slices in the 3D distribution in a longitude interval that contains the brightest part of the NPS, in an attempt to get some insight into the region responsible for the emission, in agreement with certain conditions. Those figures show that at low latitudes the densest clouds are distributed at two favored distances, 100 and 200 pc, the second corresponding to the well known Aquila Rift region. Cavities do exist between those two groups, in particular at $l=25-30^\circ$ there is a quite large ($\sim$100 pc) cavity extending -30-35$^\circ$ and +30-35$^\circ$. From the  3D figures of Lallement et al (2013), this cavity is squeezed and irregular, with tunnels opening  in several directions, in particular in the southern part. The NPS hot gas could be within this cavity between the two \textit{walls}, between 100 to 200 pc (see circle A in Fig. \ref{npsvert30}), which would  nicely correspond to the estimated distance to the Loop I center. However there is no visible tunnel linking the LB to this nearby cavity  that would correspond to the bright region ($l=20^\circ$ to $30^\circ$, $b=10^\circ$ to $30^\circ$). Moreover, surprisingly there are no nearby ($\lesssim$ 200 pc) thick clouds located below $b \sim +10^\circ$ that could explain the full disappearance of emission below this latitude that can be seen in Fig \ref{corr}. More precisely, between longitudes $l=0^\circ$ to $30^\circ$ the opacity below $b \sim +10^\circ$ and from the Sun to 100-150 pc is smaller or equivalent to the opacity at higher latitude, say between $b \sim +10^\circ$ and $b \sim +20^\circ$. This suggests that if the NPS disappearance below $b \sim +10^\circ$ is due to an increased extinction (case 1), then the X-ray source is located beyond 200 pc. In the same longitude interval the opacity starts to decrease only above $b \sim +20^\circ$.

Within the assumption of a close, emission bounded NPS (i.e. the discontinuity is not due to extinction, but to the absence of emission below this latitude the case 2), there is no obvious large nearby cavity matching the bright NPS region and located above +10$^\circ$. A potential, less obvious source shown as circle B in Figure (see Fig.\ref{npsvert30}), is a small cavity at $(l,b)=(20^\circ,18^\circ)$ and centered at about 120-130 pc. The \textit{wall} in front of this cavity is thin and its absorption would not very much influence the X-ray pattern. In this case the NPS would be a northern feature, with no counterpart in the south. This potential solution has the advantage of potentially explaining HI shells at high latitudes as extensions of this cavity. However, we are not aware of any OB associations at such latitudes which could potentially have given rise to such a cavity. Also, as we will see in the next section, absorbing columns derived from X-ray spectra do not match the opacity of the clouds in front of the cavity.
Thus, contrary to the other 5 cases mentioned above, here we could not find the same type of obvious correspondence between the X-ray pattern and the 3D distribution, if if distances are restricted to less than 200 pc.

\subsection{Inferences from maps and previous NPS spectra}

\begin{figure*}[htbp]
\centering
\includegraphics[width=0.5\linewidth]{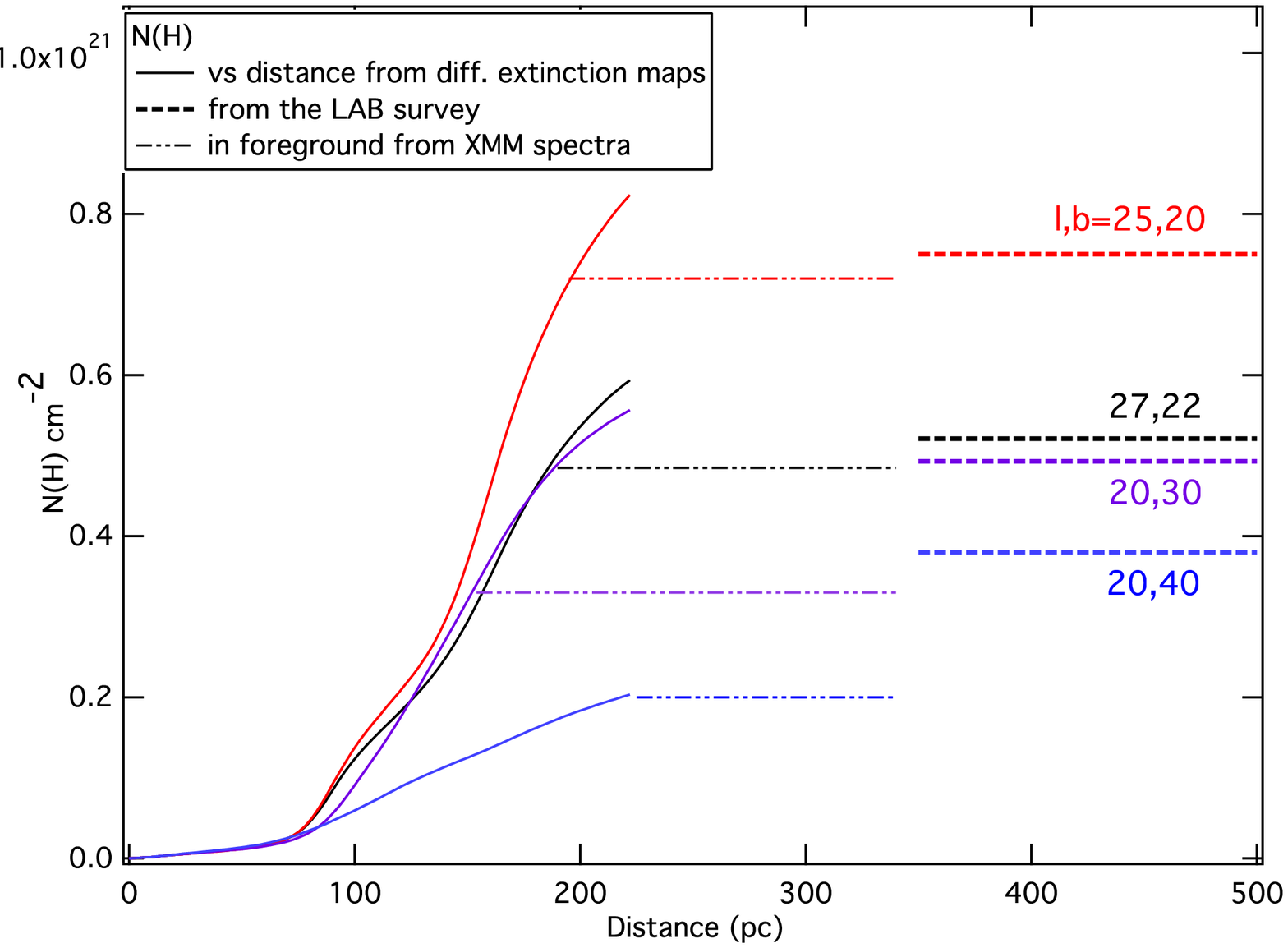}
\caption{Radial distance evolution of the opacity based on the inverted 3D maps, in the four NPS directions observed
by \cite{willi03} and \cite{miller08}  (see text). The opacities are here transformed to N(H) through the classical relation N(H)=5e21 x E(B-V). Also shown are the absorbing columns \cite{willi03} and \cite{miller08} deduced from the spectral modeling (dot-dashed lines), as well as the line-of-sight integrated HI columns from the LAB survey (dashed lines).}
\label{willmiller}%
\end{figure*}

\cite{willi03} and \cite{miller08}  have analyzed in detail XMM-Newton and Suzaku spectra towards different directions within the NPS. An output of their spectral modeling is the estimated N(H) absorbing column towards the emitting region. We have used our 3D distribution to integrate the differential opacity along the four directions they have studied. We compare in Fig. \ref{willmiller} the radial opacity profiles, converted into hydrogen column profiles using N$_{H}$= 5 10$^{21}$ E(B-V) cm$^{-2}$, with their estimated absorbing columns on one hand, and the total HI column deduced from 21 cm data \citep{kalberla05} on the other hand. 
We caution here again that the map resolution is of the order of 20-30 pc in this volume, i.e.  there is a strong averaging of the structures that explains the smooth increases in opacity. Another characteristic of the inversion must also be mentioned here, namely the use of a default distribution, a distribution that prevails at locations where there are no constraints from the reddening database. Such a default distribution has no influence on the location of the mapped clouds, but when computing opacity integrals there maybe a non negligible influence of this distribution, especially outside the Plane and at large distances  where there are fewer targets. In the maps we use here the default distribution is homogeneous and depends only on the distance from the Plane with a scale height of 200 pc. The effect of this distribution explains the smooth radial  increase of the opacity seen in Fig \ref{willmiller}  beyond about 200 pc, where targets stars are under-sampled.

With this caveat in mind, it is interesting to compare the dust (or equivalent gas) opacity profiles with the \cite{willi03} and \cite{miller08} absorbing columns, as well as with the total HI columns from 21 cm measurements. We see that for the two sight-lines at $b=20$ and $b=22^\circ$ (red and black) the absorbing columns deduced from the X-ray spectral analysis are reached at about 180 and 185 pc, and are on the order of the total (21 cm) columns. This suggests that the inner edges of the X-ray bright cavity is located beyond $\sim $180 pc. At higher latitude the absorbing columns are reached at about 150 and 220 pc respectively, and there is a significant column of gas beyond the emitting area according to the total HI columns. 
If we consider in a global manner the four directions,  the comparison between the foreground columns and the maps strongly suggest  a minimum distance of the order of 150 pc to the nearby boundary of the emitting gas. Using the $l=20^\circ$ and $l=25^\circ$ degrees slices in Fig. \ref{findnps} (two of the sightlines are at $l,b= (20,30)$ and $(20,40)$ and contained in the $l=+20^\circ$ slice, the two others are at $l,b= (25,20)$ and $(27,22)$ and contained or close to the $l=+25^\circ$ slice), it can be seen that this boundary is beyond the densest cloud complex appearing in the maps between $\sim $ 100 and 250 pc depending on directions. 
We have shown in Fig \ref{npsvert30} the dust distribution in the $l=30^\circ$ meridian plane, with three potential source regions for the bright NPS. The above  results are in contradiction with the  cavity centered at 150 pc (circle A in Fig \ref{npsvert30}) as the origin of the NPS emission, because the inner boundary of this cavity A is at $\sim $ 100 pc, and thus significantly closer than the 150pc limit. This is in agreement with the absence of a strong dust opacity increase below $b\sim 10^\circ$ in the foreground of the cavity, while this increase is necessary to explain the brightness pattern (see fig \ref{corr}). The cavity B is also very unlikely since it starts at $\sim $ 100 pc, and there is very little absorbing matter in front of it. The most likely region is the cavity C for which the absorbing column is close to the value deduced from the X-ray spectra, however there is no visible outer edge and the emission could be generated anywhere. We caution that our arguments based on the absorbing columns rest on our chosen relationship between the color excess and the gas column, which may be somewhat different in supernova heated regions. Still, the absence of a marked opacity increase below $b\sim 10^\circ$ in front of cavity A,  that disfavors it, does not depend on this relationship.

\subsection{Integrated dust opacity and 3/4 keV emission}

\begin{figure*}[htbp]
\centering
\includegraphics[width=0.7\linewidth]{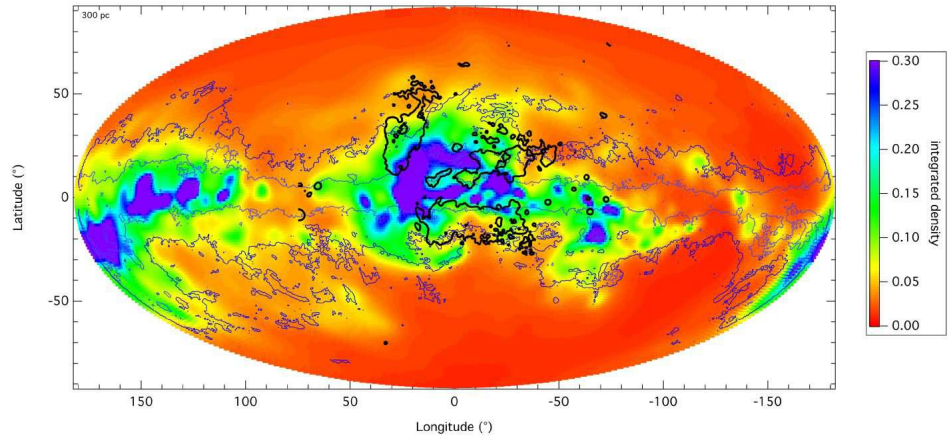}

\caption{Integrated IS dust opacity from the Sun to  300 pc (based on the differential extinction 3D maps of \cite{lallement13}).  A  0.75 keV iso-brightness contour is superimposed (black line) as well as two iso-contours drawn from the Schlegel at al  dust maps (blue and violet lines).}
\label{integrateddens}%
\end{figure*}

Based on our 3D inversion cube, we have integrated the IS opacity up to 300 pc, in order to gain a more global view of the correspondence between the nearby ISM column and the X-ray map. The 300 pc integrated opacity is shown in Fig. \ref{integrateddens}. Superimposed is an iso-brightness contour from the ROSAT 0.75 keV map. We definitely see a negative correlation  between the 0.75 keV brightness and the integrated opacity, as expected. However, there is no anti-correlation  in the region discussed above at $b=+8$ deg , $l=20,25^\circ$, where the X-ray signal has an abrupt discontinuity. 

\section{Conclusions and discussion}

We have made a first attempt to compare qualitatively and quantitatively soft X-ray background emission maps and 3D maps of the nearby IS dust. The dust maps have been recently computed by inverting $\sim$23,000 reddening measurements towards nearby stars and are particularly suited for the identification of nearby low density regions due to the large number of unattenuated target stars that form the database. Our first conclusion is that such a comparison between mapped cavities on one hand, and diffuse X-ray background on the other hand will allow to go one step further in the analysis of the hot gas X-ray emissivity in the solar neighborhood. We have illustrated the potential use of the maps in two main ways: first, we have focused on the Galactic Plane and studied the low Z ISM along with the 0.25 keV ROSAT diffuse emission pattern. A correction for heliospheric contamination of the background due to solar wind charge-exchange has been preliminarily made.  We have shown that in a number of cases the 3D structure is shedding light on the emission sources. The soft X-ray  bright regions are definitely associated to cavities, with the emission being the highest arising  from the giant cavity found along the 60(70)$^\circ$-240$^\circ$ axis, the huge super-bubble associated to the radio super-shell GSH238+08+10. A smaller and narrower cavity at $l\sim $70$^\circ$ is responsible for the second enhancement. In some cases the cloud location and the X-ray pattern are particularly well matching, e.g. an extended cloud located at about 120 pc in the direction of the CMa superbubble is shedding a strong X-ray shadow.  Finally, the study shows strong evidence for the existence of hot gas in the Local Cavity close to the Sun, a recently well debated subject. Based on the 3D dust distribution, we have computed a simplified radiative transfer model, assuming hot gas homogeneity. The data-model comparison allows to infer the average pressure in the Local Cavity, found to be on the order of 10,000 Kcm$^{-3}$, in agreement with previous studies. On the other hand, the comparison with ionized calcium absorption data confirms that the wide cavity in the third quadrant contains a significant fraction of warm ionized gas.

 In the case of the 0.75 keV maps, we find good correspondences between the existence of tunnels or cavities in the 3D maps and X-ray enhancements. At variance with these results, our search for a nearby cavity that could be at the origin of the very bright 0.75 keV emission from the North Polar Spur gave  a negative result, instead we find evidence that the NPS signal originates from hot gas beyond $\sim $ 200 pc, at a larger distance than previously inferred. The same conclusion is reached from the comparison between the dust distribution and  foreground absorbing columns deduced from NPS X-ray spectra. Since we can only derive a lower limit for the NPS source region, a much larger distance,  including up to the Galactic center, is not precluded.

More detailed 3D maps are required to improve the type of diagnostics on the X-ray emitting regions we have presented here, in particular in the case of the NPS for which there is no definite answer here. Hopefully increasing numbers of target stars will allow to built such maps and reveal more detailed IS structures. Indeed, numerous  spectra and subsequently large datasets of  individual IS absorption and extinction measurements are expected from spectroscopic surveys with  Multi-Object spectrographs. In parallel, precise parallactic distances will hopefully be provided by the ESA cornerstone GAIA mission.


\begin{acknowledgements}
We thank our anonymous referee for the numerous constructive comments on the manuscript. They resulted in a significant improvement of the article.

\end{acknowledgements}

\end{document}